\def\beg{\begin{equation}}
\def\eeq{\end{equation}}
\documentstyle[12pt]{article}
\begin{document}
\baselineskip18pt
\begin{center}
{\Large{\bf Theory of quantum Hall effect and high Landau levels}}
\vskip0.35cm
 Keshav N. Shrivastava\\
{\it School of Physics, University of Hyderabad, \\
Hyderabad  500 046, India}
\end{center}
\vskip0.5cm
The angular momentum model which couples the spin and charge is
discussed as a possible theory of the quantum Hall effect. The
high Landau level filling fractions 5/2, 7/3 and 8/3 are
understood by this model. It is found that 7/3 and 8/3 are the
particle-hole conjugates and 5/2 arises due to a limiting level
at 1/2 with Landau level number $n=5$ which makes the fraction
as 5/2.
\newpage
Recently, Eisenstein et al$^1$ have found that at large values
of the Landau level quantum number, $n$, there is much less
structure in the diagonal resistivity as a function of field
than at low values of $n$. It was pointed out by Lilly et al$^2$
that the rich structure of fractions of the fractionally
quantized Hall effect, FQHE, found at small values of $n$ is
vertually absent for large values of $n$; only fragile and
poorly understood states at Landau filling fractions, $\nu=7/3,
5/2$ and 8/3 are seen in the best samples. In a recent
letter$^3$ we have shown that our earlier representation$^4$ of
the quantum Hall states works very well for the understanding of
the particle-hole symmetry of the quantum Hall states. In
particular we are able to understand the measured equivalence in
the $g$ factor and effective masses of the quasiparticles. We
have shown that 4/5 and 6/5 have equal masses because they are
particle-hole conjugates. Yeh et al$^5$ also find that the ratio
of the effective mass to the square root of the perpendicular
magnetic field is equal for some of the fractions in agreement
with our calculations.$^3$

In the present letter, we interpret the fractions 7/3, 5/2 and
8/3 seen in the measurement carried out by Eisenstein et al$^1$.
We report that 7/3 and 8/3 are particle-hole conjugates and 5/2
is the $n=5$ states of the level at 1/2.

Our earlier model$^4$ has several features which are in
agreement with experimental measurements. The fractions
predicted by us are the same as those experimentally found by
Eisenstein and Stormer$^6$. The fractions occur in two groups.
One of the groups belongs to spin +1/2 and another to -1/2.
Therefore, the groups represent time reversed states. The
grouping of fractions in our theory is the same as
experimentally found in the measurement of Eisenstein and
Stormer$^6$. Our model also gives $\nu=1/2$ for very large
values of $l$ and hence there is a limiting value at $n/2$ where
$n$ is the Landau level quantum number. In our model one of the
series is,
\beg
\nu = {l\over(2l+1)}
\eeq
which predicts one group of fractions, 0, 1/3, 2/5, 3/7, 4/9,
5/11, etc. which are also observed by Willett et al$^7$. Another
group of fractions is predicted$^4$ by the expression,
\beg
\nu = {(l+1)\over(2l+1)}
\eeq
which are 1, 2/3, 3/5, 4/7, etc. in complete agreement with the
experimental measurement$^{6,7}$. When $l=\infty$ both the above
series approach 1/2 except that one series approaches from the
right hand side and the other from the left hand side exactly as
observed$^6$. The left and the right side approaches arise from
the Kramers conjugate states and the predicted approach is
exactly as observed$^6$. Since the limit is involved there is a
Fermi surface at 1/2. However, we can shift the Fermi surface to
higher values when higher Landau levels are occupied. The
fraction 1/2 becomes $n/2$ with $n$ as the Landau level quantum
number. The 1/2, 3/2, 5/2, 7/2, 9/2, etc. become allowed. This
predicted feature with an odd numerator with 2 in the
denominator is also exactly as observed. Thus for $n=1$, we have
two series one merging from left while the other merging from
right at 1/2 and the same picture is repeated for different
values of $n$. The entire pattern of pairwise series is observed
eactly as predicted. Many fractions with 2 in the denominators
have been observed by Lilly et al$^8$.

In the C.F. model$^9$ the spin and charge are decoupled but one
of the series is the same as our series. Another series in the
C.F. model is,
\beg
\nu_{CF} = {m\over(2m-1)}\,\,\,.
\eeq
We point out that the above series is the same as in eq.(2)
which is given in the earlier dated theory of ref. [4]. We
equate (2) and (3) so that,
\beg
 {(l+1)\over(2l+1)} = {m\over(2m-1)}
\eeq
which can be simplified to yield $m=l+1$. Therefore, the series
of (3) of C.F. is identical to that of ref. [4] except for the
shift of the integer by one. However, in the theory of ref. [4]
the spin and charge are coupled whereas in the C.F. model these
are decoupled. In the C.F. theory 1/3 and 2/3 are treated at
par. Their spin can have any value, up or down. On the other
hand in the theory$^4$ 1/3 has spin down and 2/3 has spin up.
The polarization experiment$^{10}$ demands $\gamma_e\sim1$ for
1/3, 0.5 for 2/3 and 0.3 for 3/5. This means that 1/3 is in one
group with spin down  whereas 2/3 and 3/5 are in another group
with spin up. This feature agrees better with our model of ref.
[4] than with C.F. model. It also means that the experiment
discreminates between 1/3 and 2/3 but the C.F. model does not.
Our model of ref. [4] discreminates 1/3 and 2/3 by virtue of
spin, being up in one case and down in another. The need for a
term which has spin raising and lowering operators along with a
function of orientation of the magnetic field is also clear. The
effective charge in the theory$^4$ is $(1/2)g_\pm e=\nu e$ and
the gap is $\omega_c=\nu B/mc$ but there is spin-charge coupling
as 1/3 has spin down and 2/3 has spin up. Thus charge and spin
are coupled in the theory of ref. 4 as they should be but in the
C.F. model spin and charge are decoupled. According to one
suggestion$^{11}$ spin and charge are decoupled but in our
theory fixing the value of the spin, such as 1/2, automatically
fixes the charge. Spin-charge decoupling can be obtained if $l$
and $s$ belong to two different particles. According to a recent
study$^{12-14}$ spin and charge are coupled so that as the spin
tilts, charge also moves and there is a phase transition in such
a system, which forms spin textures along with a charge density
wave. However, formation of charge-density waves by the symmetry
breaking in the electron-phonon interaction does not depend on
the spin. In the ordinary electron-phonon interaction the
electrons are scattered without change in spin.

The even denominators are indeed difficult to obtain. However,
it should be noted that the first one to be discovered was 1/3
and later on other denominators were seen. For a long time only
odd denominators were reported which shows that even
denominators are weak. For some time odd denominator rule was
established but such a rule is not correct because even
denominators occur as well as even numerators with odd
denominators are found. Later on even denominators were
reported. Therefore, it is observed that the fractional quantum
Hall effect plateaus corresponding to even denominators are weak
compared with those of odd denominators. There is a limiting
value of the series $l/(2l+1)$ which for $l\gg1$, gives 1/2
which has an even denominator. Because of the limit there is a
Fermi surface at 1/2. However, there are other even denominators
which are not linked with a Fermi surface at 1/2. Besides, when
higher Landau levels are populated the Fermi surface will have
to be shifted from 1/2 to $n/2$. Thus there are two types of
even denominators, type-I are those which are limits of the
series which give odd denominators and the type-II which are not
the limiting values. When $\nu$ is a fraction then $n\nu$ is
also allowed. Therefore, several even denominators can be
generated from 1/2 which include 1/2, 3/2, 5/2, 7/2, etc.
However, there are other types of even denominators which
require two-quasiparticle states or bound states of two quantum
Hall states. We make an effort to combine a series of the type
$l/(2l+1)$ with another series $(m+1)/(2m+1)$ in which both are
the angular momenta series of ref. 4, such that an even
denominator results,
\beg
\left[{l\over(2l+1)}\right] + {(m+1)\over(2m+1)} = {1\over(2n)}\,\,\,.
\eeq
The solution of which is,
\beg
m = {[1-2n-l(6n-2)]\over[8nl-4l+2n-2]}\,\,\,.
\eeq
For $n=2, m=-(10l+3)/(12l+2)$ which gives $l=0, m=-3/2, n=2$ and
hence $1/2n=1/4$, so that even denominators can be generated if
$m$ can become a negative fraction. For $l=1$, $m=-13/14, n=2$
and $1/2n=1/4$ is also giving an even denominator of 4 but $m$
has become a negative fraction. Thus we can obtain even
denominators for fractional angular momentum. In the Pauli spin
matrices, if 1/2 is replaced by some other fraction, the
commutators remain unchanged and it is acceptable to have values
other than 1/2 without any sacrifize of quantum mechanical
rules. The occurrence of even denominators may be anisotropic
and hence may depend on the rotation of matrices or on the
angle, which the z-axis makes with the direction of the external
magnetic field. 

The cyclotron frequency is $\omega_c=(1/2)geB/mc=\nu eB/mc$.
Therefore, we can replace the charge by an effective charge,
$e_{eff}=(1/2)ge=\nu e$. Alternatively, we can replace the field
by an effective field, $B_{eff}(1/2)geB=\nu B$ so that the field
is very much reduced. In the case of $\nu=1/3$, the field
experienced is $(1/3)B$ which is reduced from $B$.

The series (1) and (2) above can be used to explain the high
Landau levels easily. Eisenstein et al$^1$ have found that at
higher values of the Landau level quantum number, $n$, the
number of fractions observed is much less than at the lowest
Landau level. At the magnetic field of 4 to 5 Tesla only a small
number of fractions are observed, the strongest ones are at 8/3,
5/2 and 7/3. Since there is a charge versus Landau level quantum
number degeneracy, it is not possible to distinguish large
charge from a large Landau level quantum number. In the angular
momentum (1) series, $l/(2l+1)$ is the particle hole conjugate of
$(l+1)/(2l+1)$. For $l=7$ two values, 7/15 and 8/15 are
predicted and $l=\infty$ value is 1/2. When the same particle
occurs in different levels its charge remains unchanged. We can
multiply the values by $n=5$ so that the predicted values of
1/2, 7/15 and 8/15 become 5/2, 7/3 and 8/3. These predicted
values are exactly the same as those observed experimentally by
Eisenstein et al. Thus 7/3 is the particle-hole conjugate of 8/3
as seen below for $n=5$:
\vskip0.5cm
\begin{tabular}{ccccc}
\hline
$l$ & $l/(2l+1)$ & $(l+1)/(2l+1)$ & $nl/(2l+1)$ & $n(l+1)/(2l+1)$\\
$\infty$ & 1/2 & 1/2 & 5/2 & 5/2\\
7 & 7/15 & 8/15 & 7/3 & 8/3\\
\hline
\end{tabular}
\vskip0.5cm
\noindent Thus the angular momenta series (1) and (2) given by
ref. 4 explain the quantum Hall effect correctly.
\vskip0.35cm
\noindent {\bf Conclusions.}
\vskip0.35cm
In conclusion, we have shown that the correct series of the
quantum Hall effect is derivable from angular momentum. The
particle-hole symmetry recently observed in quantum Hall effect
is understable from the angular momentum and even denominators
are predicted correctly. We are also able to understand the
high Landau levels$^{15-16}$.

\newpage
\noindent{\bf References.}
\begin{enumerate}
\item J.P. Eisenstein, M.P. Lilly, K.B. Cooper, L.N. Pfeiffer
and K.W. West, condmat/9909238; 13$^th$ International
conference on the electronic properties of two dimensional
systems, Ottawa, Canada, 1999; Physica E, {\bf6}, 29 (2000).
\item The 7/3, 5/2 and 8/3 are also mentioned in M.P. Lilly,
K.B. Cooper, J.P. Eisenstein, L.N. Pfeiffer and K.W. West, Phys.
Rev. Lett. {\bf82}, 394 (1999).
\item K.N. Shrivastava, Mod. Phys. Lett. B {\bf13}, 1087 (1999).
\item K.N. Shrivastava, Phys. Lett. A{\bf113}, 435 (1986);
{\bf115}, 459(E) (1986).
\item A.S. Yeh, H.L. St\"ormer, D.C. Tsui, L.N. Pfeiffer, K.W.
Baldwin and K.W. West, Phys. Rev. Lett. {\bf82}, 592 (1999).
\item J.P. Eisenstein and H.L. St\"ormer, Science, {\bf248}, 1510 (1990).
\item R. Willett et al, Phys. Rev. Lett. {\bf59}, 1776 (1987).
\item M.P. Lilly, K.B. Cooper, J.P. Eisenstein, L.N. Pfeiffer and
K.W. West, Phys. Rev. Lett. {\bf83}, 824 (1999).
\item J.K. Jain, Phys. Rev. Lett. {\bf63}, 199 (1989).
\item I.V. Kukushkin, K. von Klitzing and K. Eberl, Phys. Rev.
Lett. {\bf82}, 3665 (1999).
\item C. Kim et al, Phys. Rev. Lett, {\bf77}, 4054 (1996).
\item S.M. Reimann, M. Koskinen, S. Viefers, M. Manninen and B.
Mottelson, Nordita-199/36CM, cond-mat/9908208.
\item A. Karlhede and K. Lejnell, Physica E{\bf1}, 41 (1997).
\item M. Franco and L. Brey, Phys. Rev. {\bf56}, 10383 (1997).
\item K. N. Shrivastava, Mod. Phys. Lett. {\bf14},1009 (2001).
\item K. N. Shrivastava, CERN LIBRARIES, GENEVA SCAN-0103007,(2000?).
\end{enumerate}
\end{document}